\documentclass{elsart1p}

\usepackage{graphics}
\usepackage{epsfig}

\usepackage{amssymb}

\begin{document}

\begin{frontmatter}


\title{Lepton-pair production in nuclear collisions - \\
past, present, future}


\author{Hans J. Specht}

\address{Physikalisches Institut, Universit\"{a}t Heidelberg, Heidelberg, Germany }

%\ead{specht@physi.uni-heidelberg.de}

\begin{abstract}
% Text of abstract
The key results on lepton-pair production in ultra-relativistic
nuclear collisions are shortly reviewed, starting at the roots of $pp$
collisions in the seventies, and ending at the perspectives of the
colliders RHIC and LHC. The presence is dominated by the recent
precision results from NA60 at the CERN SPS, culminating in the first
measurement of the in-medium $\rho$ spectral function and the
transverse flow of the associated thermal radiation. The seeming
cut-off of the flow above the $\rho$ may well be the first direct hint
for thermal radiation of partonic origin in nuclear collisions. The
major milestones in the theoretical developments are also covered.

\end{abstract}

\end{frontmatter}

% main text
\section{The past: from $pp$ to the first results in $AA$ collisions}
\label{past}

The interest in continuum lepton-pair production in high-energy
collisions dates back to the seventies, triggered by the detection of
the Drell-Yan process~\cite{Drell:1970wh} and the $J/\psi$. The
latter, in particular, sharpened the attention to anything which might
still have escaped detection, and a flood of new experimental findings
on lepton pairs appeared, both for low masses (LMR, M$<$1 GeV) and for
intermediate masses (IMR, 1$<$M$<$2.5 GeV). The results were usually
compared to expectations from an ``hadron-decay cocktail'', containing
all contributions known at that time. An excess of single leptons and
lepton pairs above the known sources was indeed found, coined
``anomalous'' pairs, and created great excitement. A review of the
situation in the LMR region as of 1984 is contained
in~\cite{Specht:1986kd}. Unfortunately, the results did not survive
critical reassessment in later years, and they were finally recognized
by Helios-1~\cite{Akesson:1994mb} and, with higher precision, by
CERES~\cite{Agakishiev:1998mw} as due to a severe underestimate of the
contribution from $\eta$ Dalitz decays. Only one result, obtained at
the ISR at $\sqrt{s}$=63 GeV~\cite{Akesson:1984sf}, survived as
non-trivial up to today. In the IMR region, some excess-pair
production was also suspected for a long time, due to insufficient
knowledge of the contribution from open charm decays on top of
Drell-Yan. Any significant anomaly in this region was only ruled out
much later, see e.g.~\cite{Abreu:2000nj}.

Ironically, these dubious $pp$ results led already in the late
seventies to two seminal theoretical papers, which had an enormous
influence on the nascent field of high-energy $AA$ collisions. Bjorken
and Weisberg~\cite{Bjorken:1975dk} were the first to propose partons
{\it produced} in the collision to be a potential further source of
continuum lepton pairs, beyond the intrinsic partons in the collision
partners responsible for Drell-Yan; they estimated the resulting
excess above the latter to be a factor of 10-100 in the LMR
region. Shuryak~\cite{Shuryak:1978ij} proposed the production of
deconfined partons in thermal equilibrium during the collision and
phrased the terms ``Quark Gluon Plasma'' for the created medium and
``Thermal Radiation'' for the emitted lepton pairs in the IMR
region~\cite{Branson:1977cj}.

The first systematic discussion, including both particle and nuclear
physicist, on the experimental and theoretical aspect of QGP formation
in ultra-relativistic nuclear collisions took place at the Quark
Matter Conference 1982 in Bielefeld~\cite{QM:1982}. The basic
instrumental elements of the first-generation experiments at the CERN
SPS as well as the basic theoretical ideas on all observables were
addressed. The principal conclusions for lepton pairs were as
follows. (i) The physics of dileptons (virtual photons) may be both
more rich and more rigorous than that of real photons, due to the
existence of two independent variables instead of one ($M$, $p_{T}$
vs. $p_{T}$), and due to the simpler lowest-order rates
($\propto\alpha^{2}_{em}$ vs. $\alpha_{em}$$\cdot$$\alpha_{s}$). (ii)
Thermal dilepton production in the LMR region may be dominantly
hadronic, mediated by the broad vector meson $\rho$ (770) in the form
$\pi^{+}\pi^{-} \rightarrow \rho \rightarrow \l^{+}l^{-}$; due to its
short lifetime of only 1.3 fm, the observation of a ``melting''
(broadening) and/or mass shift may serve as a prime probe of {\it
chiral symmetry restoration}~\cite{Pisarski:mq}. (iii) Thermal
dilepton production in the IMR region may be dominantly partonic,
mediated in the form $q\bar{q} \rightarrow l^{+}l^{-}$, and may serve
as a prime probe of {\it deconfinement} (the idea of $J/\psi$
suppression was not yet born). Classical theoretical papers on
continuum lepton pairs with a broad view appeared soon
after~\cite{McLerran:1984ay,Kajantie:1986}.

\begin{figure*}[t!]
%\centering
\hspace*{0.3cm} \includegraphics*[width=6.0cm, height=5.8cm]{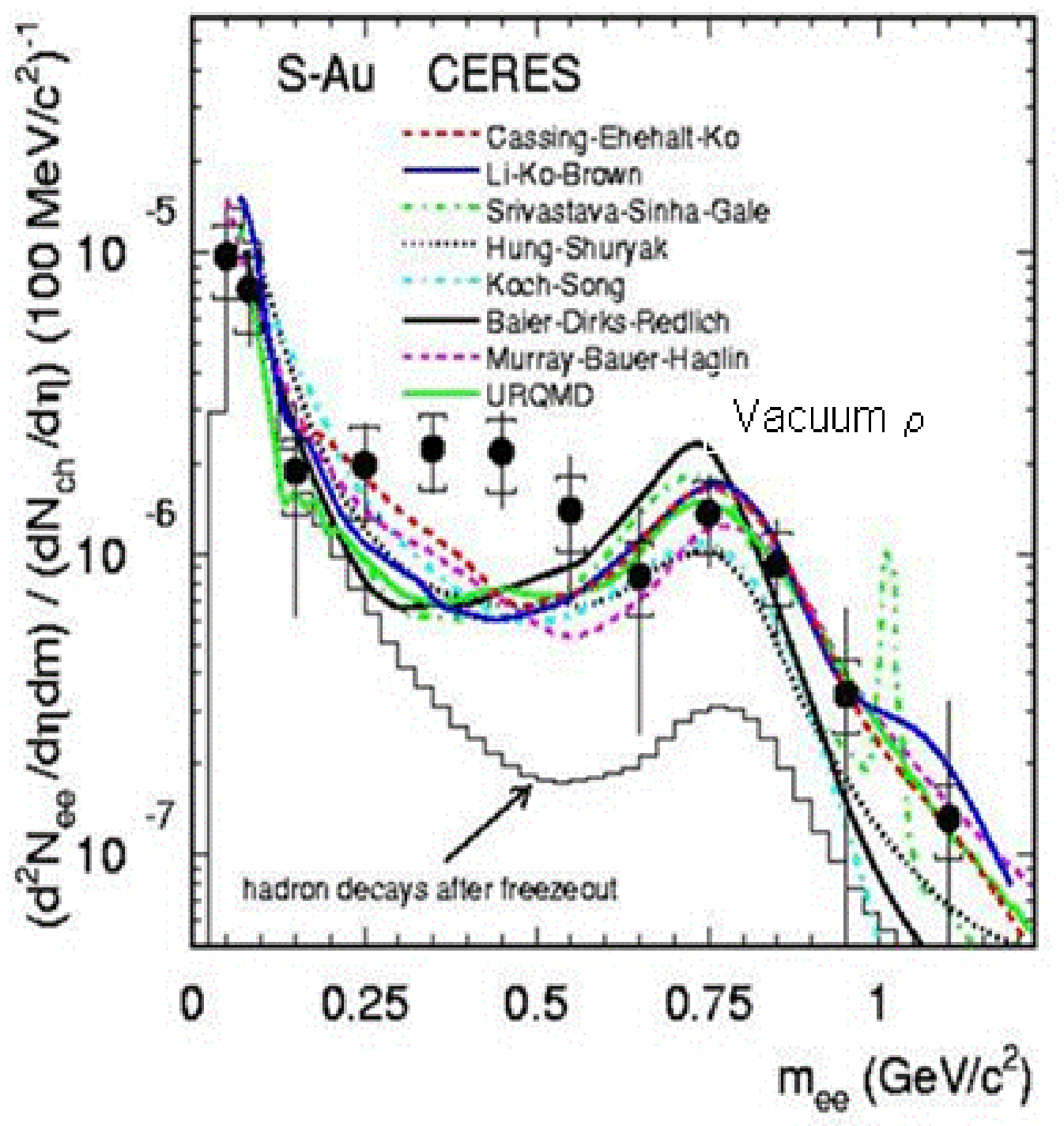} \hspace*{0.4cm}
\includegraphics*[width=6.0cm, height=6.0cm,clip=, bb= 0 37 520 659]{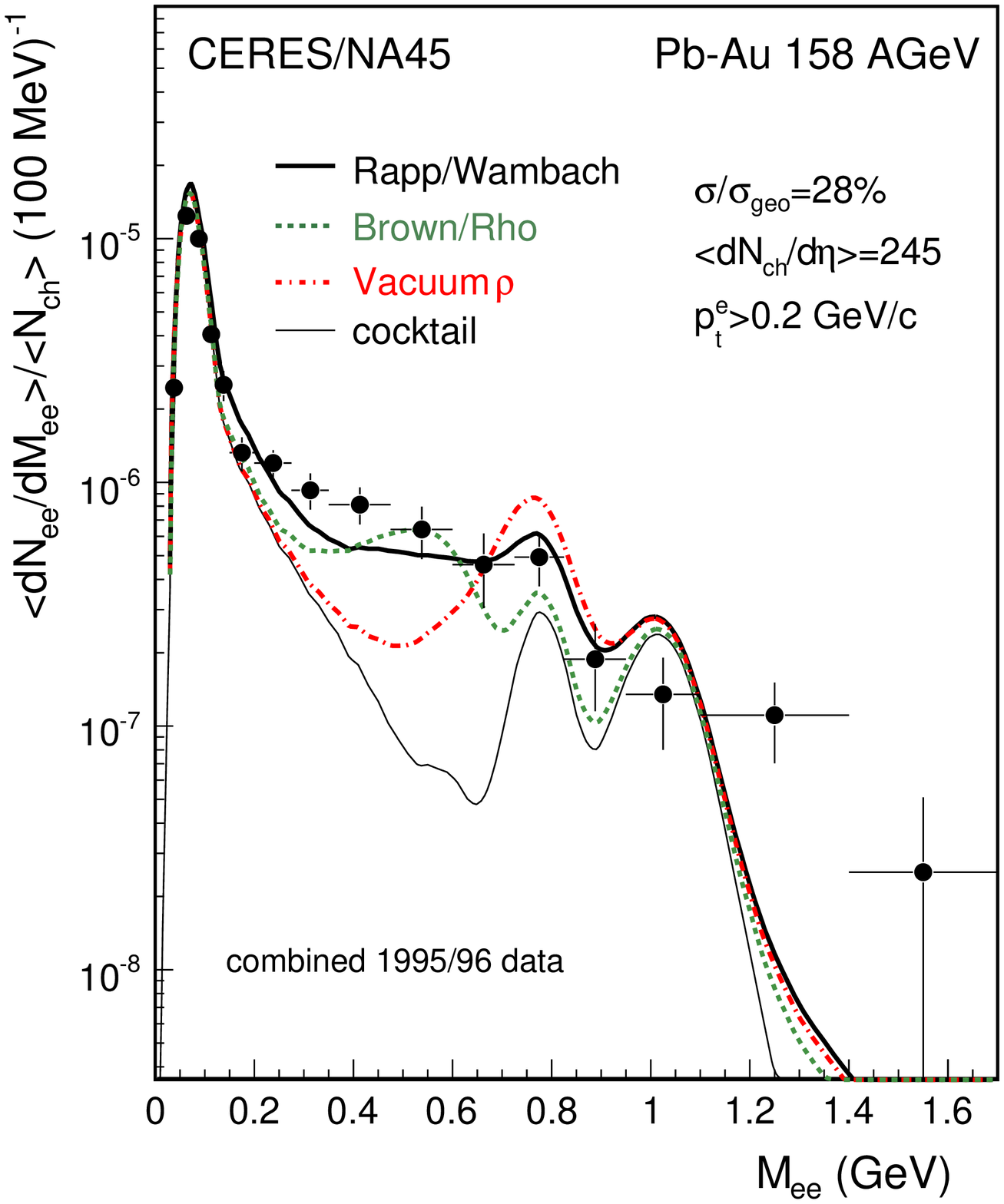}
\caption{Comparison of CERES
results~\cite{Agakichiev:1995xb,Agakichiev:1997au} to theoretical
modeling (see text).}
\label{fig1}
\end{figure*}

The first generation of SPS experiments sensitive to continuum lepton
pairs, Helios-2 and NA38, found one
anomaly~\cite{Abreu:2000nj}, but did not follow up its significance at
that time. Only in the next generation, with CERES, Helios-3 and NA50,
clear signs of new physics appeared in a broader way, 13 years or more
after the Bielefeld workshop.  The experimental results from CERES for
S-Au~\cite{Agakichiev:1995xb} in the LMR region are shown in
Fig.~\ref{fig1} (left). A large excess of pairs above the known hadron
decays is seen. This gave an enormous boost to theory, with hundreds
of publications. A small fraction of those, relying on $\rho$
production without in-medium effects, is contained in the figure. The
pole region is enhanced because of regeneration via $\pi^{+}\pi^{-}
\rightarrow \rho$ during the fireball expansion, but the bulk of the
excess residing below the pole is not at all described. Only
switching-on in-medium effects, e.g. mass shifts, based on a direct
connection to the restoration of chiral symmetry~\cite{Brown:kk}, or
broadening, based on a hadronic many-body approach~\cite{Rapp:1995zy},
leads to a satisfactory description, while not discriminating between
the two. This ambivalent situation also persisted into the Pb-beam
era, as illustrated in Fig.~\ref{fig1} (right) for the CERES 1995/96
data~\cite{Agakichiev:1997au} (and still valid for the 2000
data~\cite{Miskowiec:2005dn}): the main two
scenarios~\cite{Rapp:1999ej,Brown:2001nh} fit the data equally well,
and the true in-medium properties of the $\rho$ could unfortunately
not be clarified, due to insufficient data quality.

\begin{figure}[h!]
%\centering
\hspace*{0.4cm} \includegraphics*[width=0.42\textwidth]{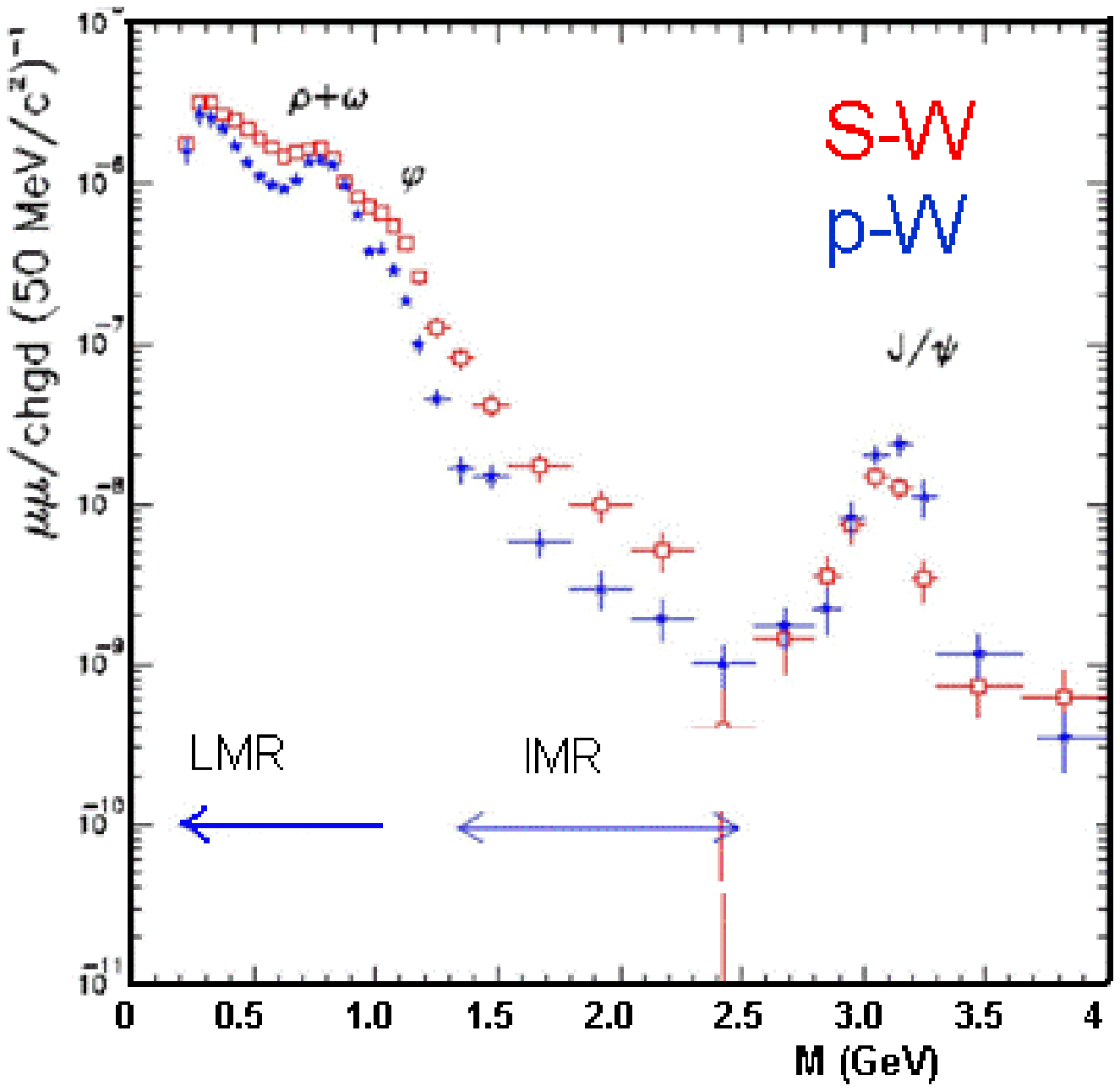} \hspace*{1cm}
\includegraphics*[width=0.42\textwidth]{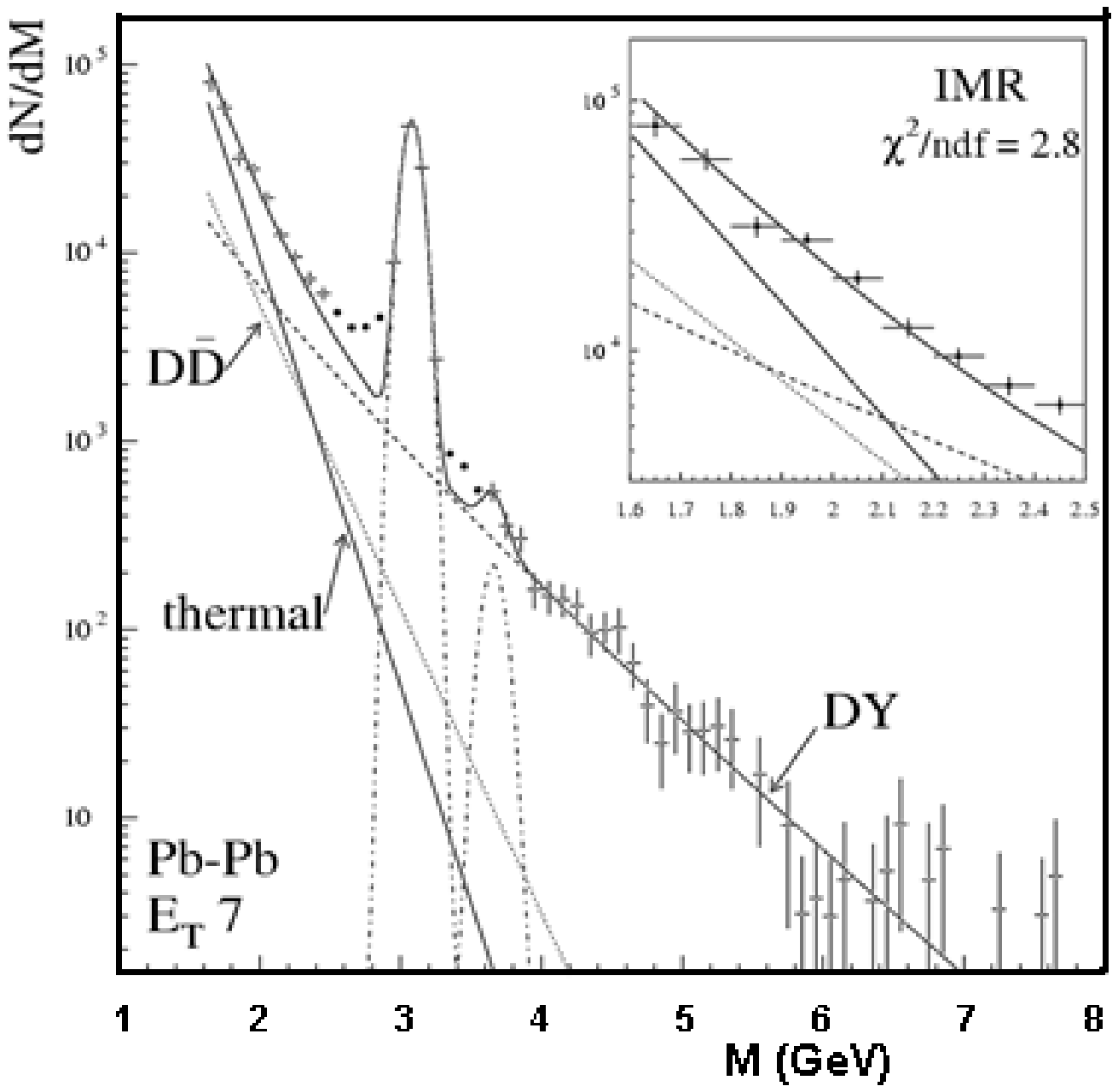}
\caption{Dilepton invariant mass spectra from Helios-3~\cite{Masera:1995ck} and NA50~\cite{Abreu:2002rm} (see text).}
\label{fig2}
\end{figure}

The excess of pairs observed by Helios-3~\cite{Masera:1995ck} for S-W
with respect to p-W is shown in Fig.~\ref{fig2} (left). It is seen
% figure 1
here to also occur in the IMR region, not only at low masses, and led
to a further important theoretical step: the recognition of the role
of chiral (V-A) mixing with possibly sizable contributions from $\pi
a_{1}$ processes~\cite{gale:nn}. A strong excess of pairs was finally
% figure 2
% end of figure 2
also reported by NA50~\cite{Abreu:2002rm}, Fig.~\ref{fig2} (right),
attributed at that time to either enhanced charm production or thermal
radiation.  The former was never followed up theoretically, while the
latter received a quantitative description in terms of hadron-parton
duality, not explicitly specifying the sources \cite{Rapp:1999zw}
(see plot). Experimentally, the role of open charm and the nature of
the thermal sources dominating the IMR region remained open.

\section{The present: results from NA60 at the CERN SPS}
\label{present}

NA60 is a third-generation experiment, built specifically to follow up
the open issues addressed in the previous section. It combines the
muon spectrometer previously used by NA50 with a novel radiation-hard
silicon pixel vertex tracker, embedded in a 2.5 $T$ dipole magnet in
the target region~\cite{Gluca:2005}. Track matching between the two
spectrometers, both in coordinate and momentum space, improves the
dimuon mass resolution by a factor of 4 relative to NA50 and allows to
distinguish prompt from decay dimuons, while the radiation hardness of
the Si tracker together with a very high readout speed allow to
maintain the high luminosity of common dimuon experiments. The
enormous jump in technology is responsible for the corresponding jump
in data quality now achieved by NA60.

% figure 3
\begin{figure*}[h!]
%\centering
\hspace*{0.6cm} \includegraphics*[width=5.4cm, height=6.0cm]{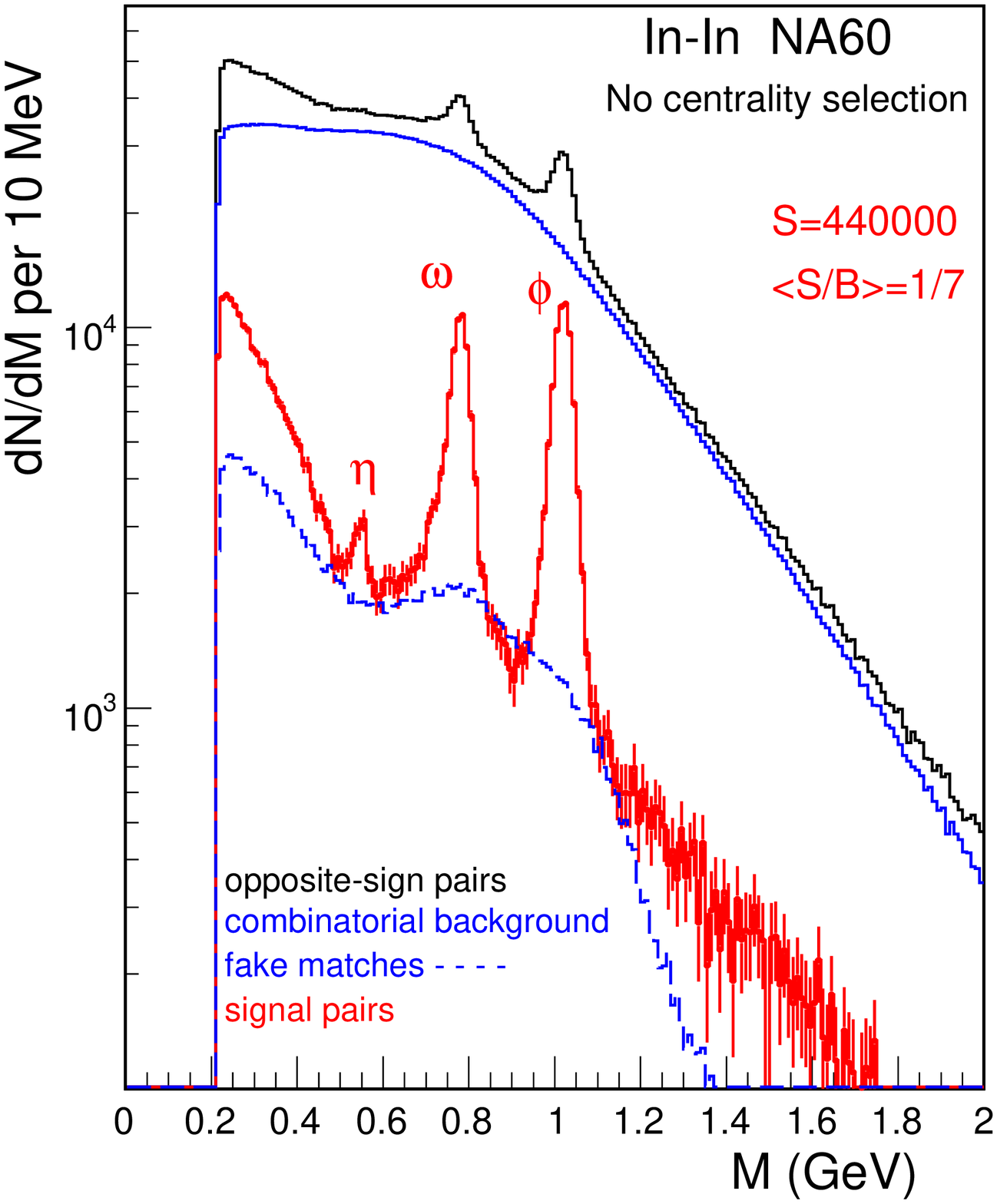}
\hspace*{0.4cm} \includegraphics*[width=5.3cm, height=6.0cm]{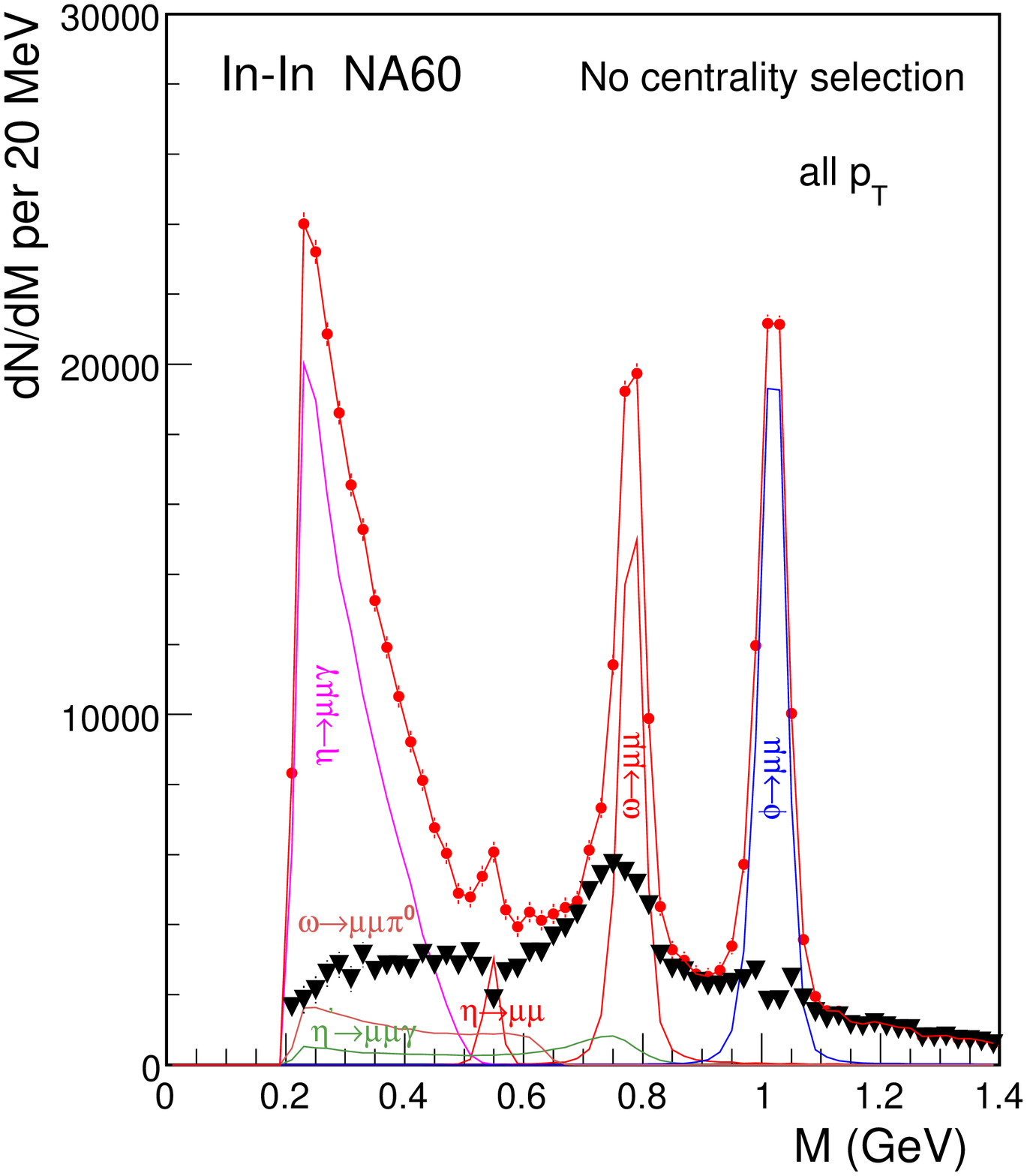}
\vspace*{-0.3cm}
\caption{Dilepton invariant mass spectra~\cite{Arnaldi:2006jq,Damjanovic:2006bd}. Right: isolation of the excess (triangles, see text).}
\label{fig3}
\end{figure*}
% end of figure 3

The results reported here were obtained from the analysis of data
taken in 2003 for In-In at 158 AGeV. Fig.~\ref{fig3} (left)
illustrates the data sample obtained in the LMR
region~\cite{Arnaldi:2006jq,Damjanovic:2006bd,Damjanovic:2007qm}. After
subtracting the combinatorial background, determined by event mixing,
and the signal fake matches between the tracks of the two
spectrometers, determined by overlay MC, the resulting net spectrum
contains about 440\,000 muon pairs. The vector mesons $\omega$ and
$\phi$ are completely resolved; the mass resolution at the $\omega$ is
20 MeV. The improvements over CERES are a factor of $>$1000 in
statistics and a factor of 2-5 in mass resolution. The centrality of
the events is tagged by determining the associated charged-particle
multiplicity density $dN_{ch}/d\eta$ from the tracks of the Si
telescope. The peripheral data (4-30) can essentially be described by
the expected electromagnetic decays of the neutral mesons, i.e. the
2-body decays of the $\eta$, $\rho$, $\omega$ and $\phi$ resonances
and the Dalitz decays of the $\eta$, $\eta^{'}$ and
$\omega$~\cite{Arnaldi:2006jq,Damjanovic:2006bd}. This is not true for
the total data as plotted in Fig.~\ref{fig3} (right), due to the
existence of a strong excess of pairs. To isolate this excess with
{\it a priori unknown characteristics} without any fits, the cocktail
of the decay sources is subtracted from the total data using {\it
local} criteria, which are solely based on the mass distribution
itself. The $\rho$ is not subtracted. The excess resulting from this
difference formation is illustrated in the same figure. In practice,
the procedure is done separately for each centrality window and, in
connection with the $p_{T}$-spectra, for each $p_{T}$ bin
(see~\cite{Arnaldi:2006jq,Damjanovic:2006bd,Damjanovic:2007qm} for
details and error discussion).

\begin{figure}[h!]
\centering
\includegraphics*[width=6.0cm, height=5.7cm,]{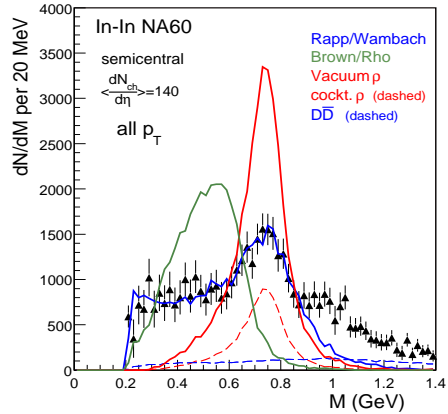}
\caption{Comparison of the NA60 excess data to theoretical model
results~\cite{Arnaldi:2006jq} (see text).}
\label{fig4}
\end{figure}

The evolution of the spectral shape of the excess
vs. centrality~\cite{Arnaldi:2006jq,Damjanovic:2006bd,Damjanovic:2007qm}
is most remarkable: a peaked structure is always seen, residing on a
broad continuum with a yield strongly increasing with centrality, but
remaining essentially centered around the position of the nominal
$\rho$ pole. These qualitative features are consistent with an
interpretation of the excess as dominantly due to $\pi^{+}\pi^{-}$
annihilation.  Fig.~\ref{fig4} shows the data for
110$<$$dN_{ch}/d\eta$$<$170 in comparison to the two main theoretical
scenarios discussed before: broadening of the $\rho$
(Rapp/Wambach~\cite{Rapp:1999ej}), and dropping mass of the $\rho$
(Brown/Rho~\cite{Brown:2001nh}), both evaluated for In-In at
$dN_{ch}/d\eta$=140 for the same fireball
% figure 4
% end of figure 4
evolution~\cite{Rapp:2005}. Since agreement between modeling and data
would imply agreement both in shape and yield, the model results are
normalized to the data in the mass interval M$<$0.9 GeV, just to be
independent of the uncertainties of the fireball evolution. The
unmodified $\rho$, also shown in Fig.~\ref{fig4} (Vacuum $\rho$), is
clearly ruled out. The dropping-mass scenario is also ruled out,
showing the much improved discrimination power of NA60. The broadening
scenario, on the other hand, fits perfectly well for M$<$0.9 GeV. It
is important to note that the data as shown have not been corrected
for the mass- and $p_{T}$-dependent acceptance of the NA60 setup,
requiring the model results to be propagated through the acceptance
filter for fair comparison to the data. By {\it pure coincidence}, as
long as no $p_{T}$ cut is applied, that filtering nearly compensates
for all the phase space factors associated with the thermal dilepton
radiation, and just leaves a mass spectrum equivalent to the spectral
function of the $\rho$, averaged over momenta and the complete
space-time evolution of the fireball, within an accuracy of about
10\%~\cite{Damjanovic:2006bd}. The flat part of the measured spectrum
may thus reflect the early history close to the QCD phase boundary,
while the narrow peak on top may be due to the late part close to
thermal freeze-out.

By now, several new sets of model descriptions have
appeared~\cite{vanHees:2006ng,Ruppert:2006hf,Dusling:2007rh}. All of
them are based on broadening of the $\rho$, and all of them describe
the data reasonably well, even in absolute terms. The fireball
evolution, quite different in these sets, takes explicit account both
of temperature and of baryon density, and the latter seems clearly
required to describe the low-mass tail of the spectral
function~\cite{vanHees:2006ng}, in accord with conclusions reached
previously by CERES from an increase of the excess at the lower beam
energy of 40 AGeV~\cite{Adamova:2002kf}. The mass region M$>$0.9 GeV
is now also addressed and again reasonably well described, in terms of
either hadronic processes~\cite{vanHees:2006ng} (4$\pi$...), or
partonic processes~\cite{Ruppert:2006hf} ($q\bar{q}$), reflecting the
traditional ambivalence of hadron-parton duality in the mass domain in
this region.

How could one distinguish? As already emphasized above, lepton pairs
are characterized by {\it two} variables, $M$ and $p_{T}$. The latter
contains not only contributions from the spectral function, but
encodes in fact the key properties of the expanding fireball,
temperature and transverse (radial) flow. In contrast to hadrons,
however, which always receive the full asymptotic flow at the moment
of decoupling, lepton pairs are continuously emitted during the
evolution, sensing small flow and high temperatures at early times,
and increasingly larger flow and smaller temperatures at later
times. Potentially therefore, the resulting space-time folding over
the temperature-flow history offers access, through the measurement of
$p_{T}$ spectra, to the emission region of the dileptons and may thus
differentiate between a hadronic and a partonic nature of the emitting
source~\cite{Kajantie:1986,Asakawa:1993kb}.

% figure 5
\begin{figure*}[h!]
%\centering
\hspace*{0.3cm} \includegraphics*[width=0.42\textwidth]{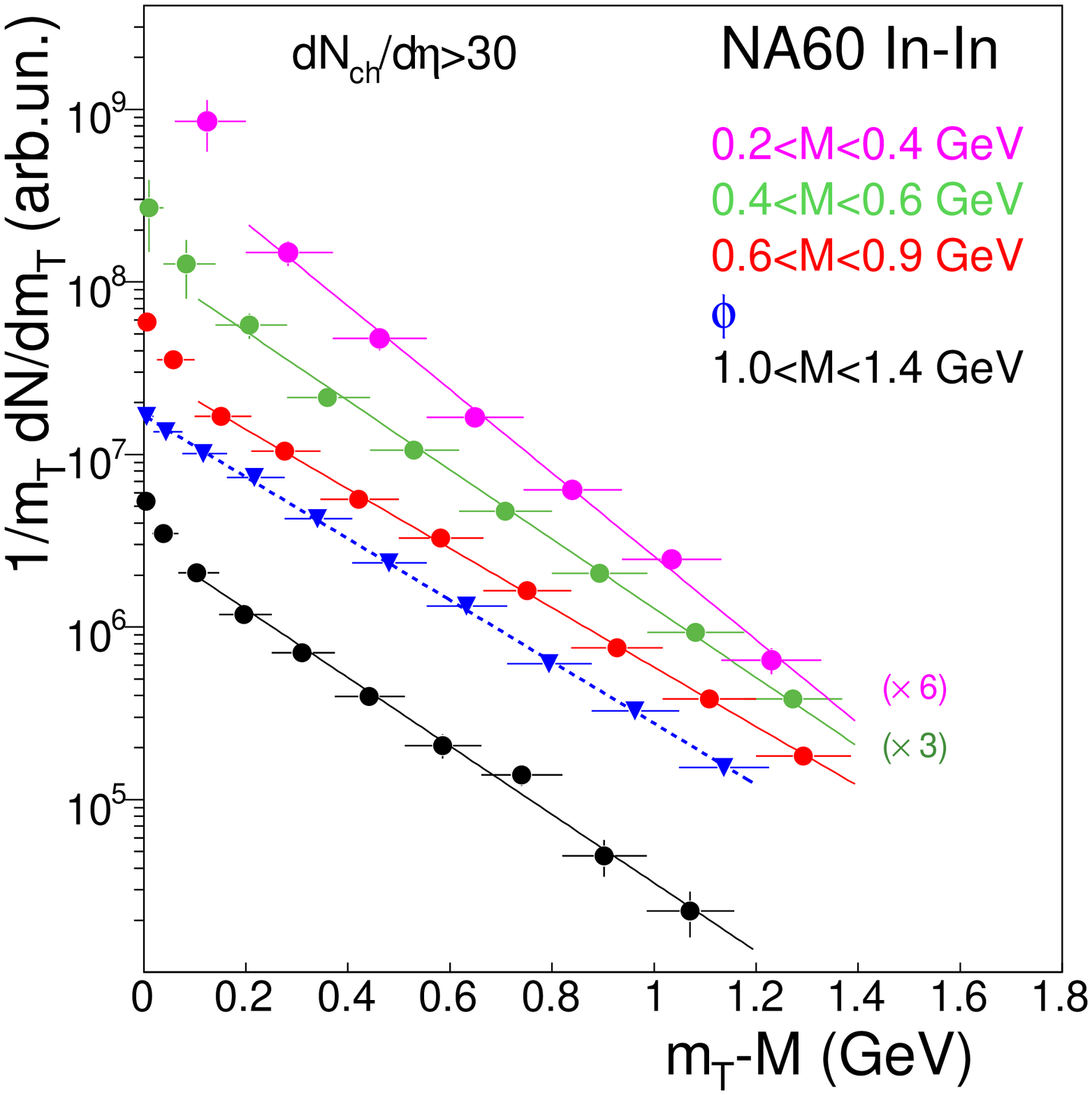}
\hspace*{0.7cm} \includegraphics*[width=0.42\textwidth]{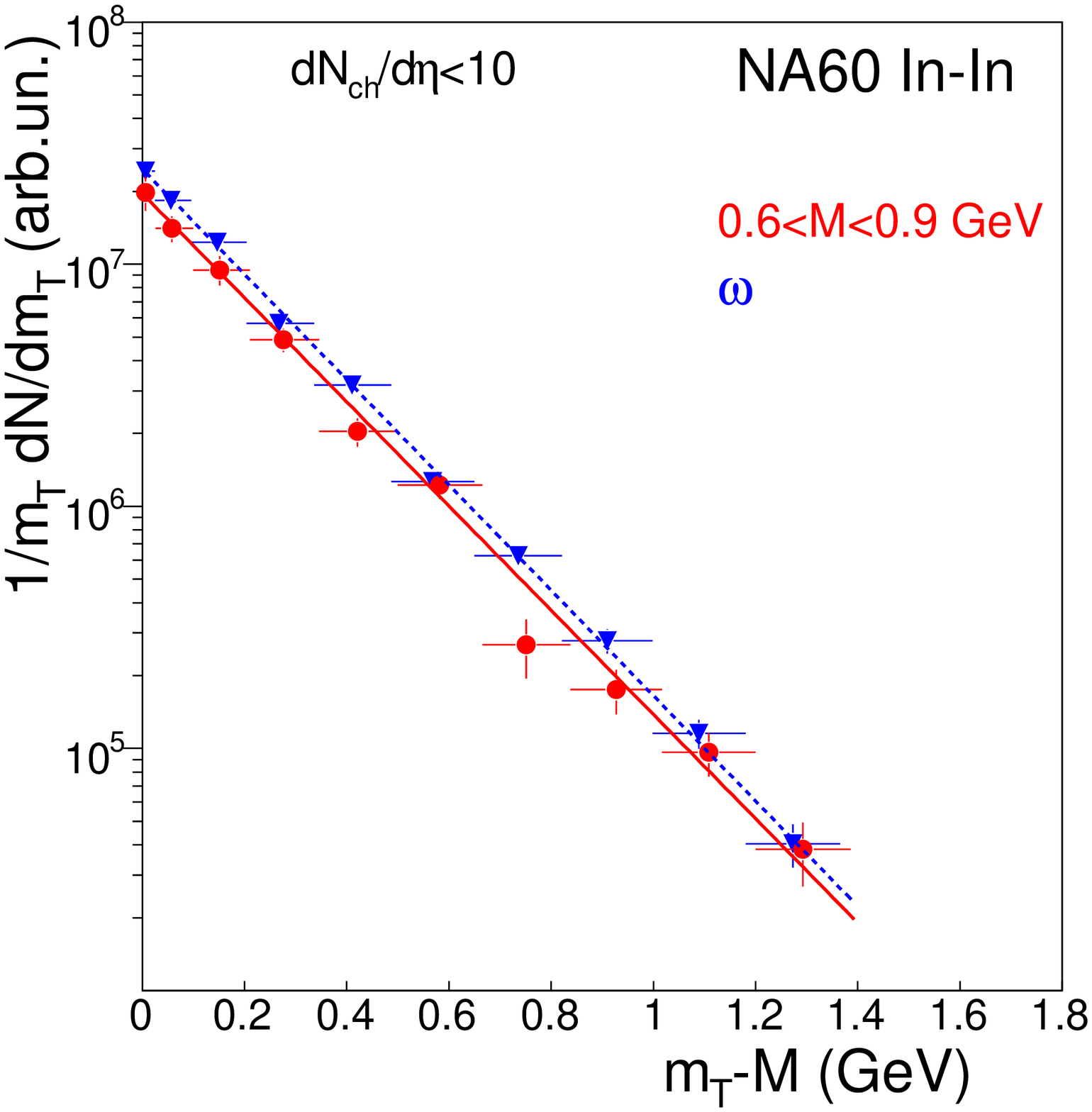}
\caption{Acceptance-corrected transverse mass spectra of the excess
(in four mass windows), the $\phi$ and the $\omega$. Left: all
centralities~\cite{Damjanovic:2007qm,Usai:2006qm,Arnaldi:prl07}. Right:
very peripheral events~\cite{Arnaldi:prl07}.}
\label{fig5}
\end{figure*}
% end of figure 5

$P_{T}$-spectra of the excess radiation in the LMR region, fully
corrected for acceptance, have recently become
available~\cite{Damjanovic:2007qm,Usai:2006qm,Arnaldi:prl07}. The
acceptance correction is done in a fine grid in the $M$-$p_{T}$ plane
to be independent of the properties of the source; details can be
found in ~\cite{Damjanovic:2007qm,Arnaldi:prl07}. Fig.~\ref{fig5}
(left) displays the centrality-integrated data in the form of
invariant $m_{T}$-spectra, where $m_{T}=(p_{T}^{2} + M^{2})^{1/2}$,
for four mass windows. At very low $m_{T}$, a steepening is observed
in all cases, contrary to the expectation for radial flow at masses
above the pion mass; the $\phi$, plotted for comparison, does not show
that (while pions themselves do). For very peripheral data, shown in
Fig.~\ref{fig5} (right), the phenomenon disappears, and the
$\rho$-like window and the $\omega$ become identical. At higher
$m_{T}$, the spectra monotonically flatten with mass, reminiscent of
radial flow, but then steepen again above the $\rho$; this striking
feature will be addressed below.

\begin{figure*}[t!]
%\centering
 \includegraphics*[width=0.44\textwidth]{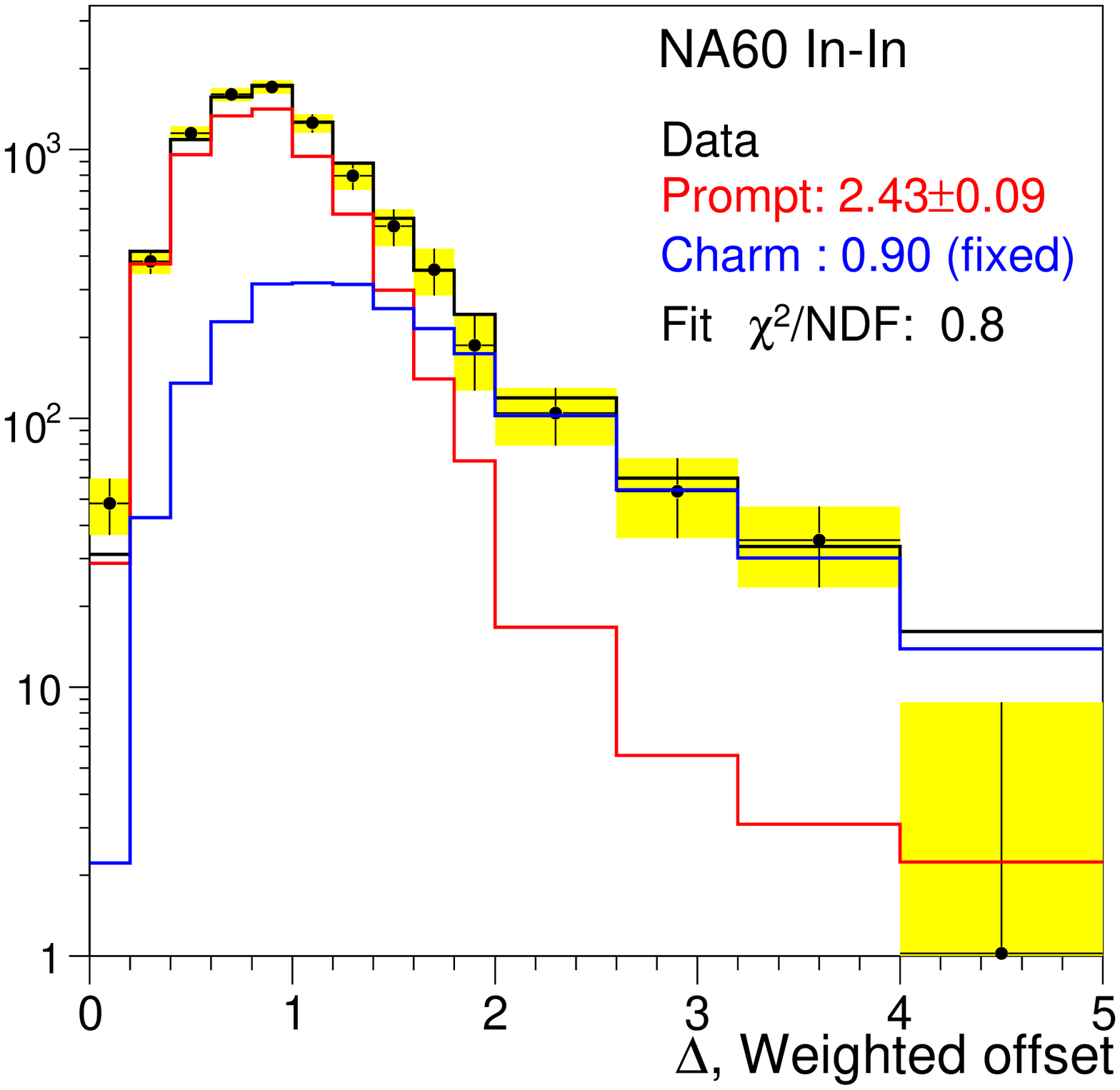}
\hspace*{0.7cm} \includegraphics*[width=0.44\textwidth]{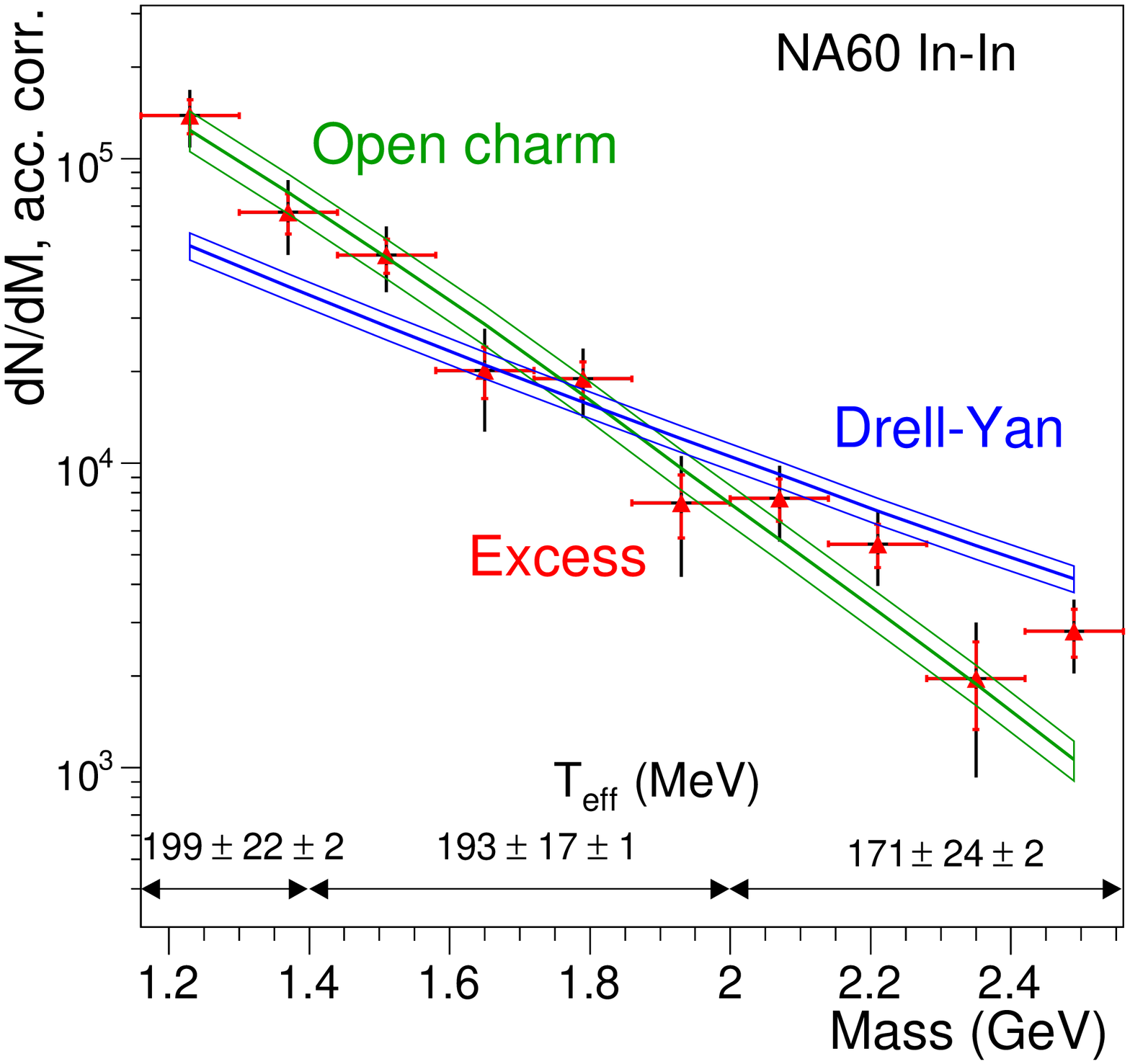}
\caption{Left: fit of the weighted offset distribution with a fixed
charm and a free prompt yield. Right: acceptance-corrected mass
spectra of Drell-Yan, open charm and the
excess~\cite{Shahoyan:2006qm}.}
\label{fig6}
\end{figure*}

The central NA60 results for the IMR
region~\cite{Usai:2006qm,Shahoyan:2006qm,Arnaldi:inpc} are illustrated
in Fig.~\ref{fig6}. By measuring the offset between the muon tracks
and the main interaction vertex, the contributions to the prompt and
to the offset part (from $D$ decays) can be disentangled. 
% figure 6
% end of figure 6
As shown in the left panel, the offset distribution is perfectly
consistent with no charm enhancement, expressed by a fraction of 0.9
of the canonical level, while the enhancement seen before by
NA50~\cite{Abreu:2002rm} and now reconfirmed by NA60 has to be solely
attributed to the prompt part, expressed by a factor of 2.4 of the
Drell-Yan level: a further milestone in experimental results. The
right panel contains the decomposition of the total data into
Drell-Yan, open charm and a prompt excess obtained as the difference
with respect to the total. The excess part is seen to exhibit the same
fall-off vs. mass as charm, steeper than DY. It increases with
centrality, and its transverse momentum spectra are also much steeper
than DY~\cite{Usai:2006qm,Shahoyan:2006qm}. The fit temperatures of
the $m_{T}$-spectra associated with 3 mass windows are indicated on
the bottom of the figure. There are all indications that this is the
thermal radiation addressed already by NA50 as one the options (see
Fig.~\ref{fig2}, right).

% figure 7
\begin{figure}[h!]
\centering
\includegraphics*[width=6.3cm, height=5.8cm]{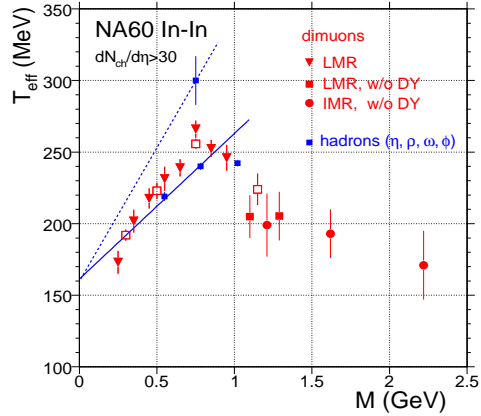}
\caption{Inverse slope parameter $T_{eff}$ vs. dimuon mass $M$ for the
combined LMR/IMR regions~\cite{Arnaldi:prl07,Arnaldi:inpc}.}
\label{fig7}
\end{figure}
% end of figure 7
The central information extracted from the $m_{T}$ spectra, expressed
as the inverse slope parameter $T_{eff}$ vs. dimuon mass, is shown in
Fig.~\ref{fig7}~\cite{Usai:2006qm,Arnaldi:inpc}. This unifies the data
from the LMR and IMR regions, including a common fit range
(0.4$<$$p_{T}$$<$1.8 GeV) and subtraction of charm throughout. The
hadron data for $\eta$, $\omega$ and $\phi$ obtained as a by-product
of the cocktail subtraction are also plotted. The parameter $T_{eff}$
is seen to rise nearly linearly with mass up to the pole position of
the $\rho$, followed by a sudden decline to a rather constant level of
190 MeV above. The excess data along the rise are quite close to the
hadron data. However, the peak of the $\rho$, residing on the broad
underlying continuum, can be interpreted as the freeze-out (vacuum)
$\rho$ without in-medium
effects~\cite{vanHees:2006ng,Ruppert:2006hf,Dusling:2007rh} and
analyzed separately by a side-window subtraction method, resulting in
a value for $T_{eff}$ of about 300 MeV (also contained in
Fig.~\ref{fig7}). The observed linear rise of $T_{eff}$ with mass of
the excess can therefore be considered as qualitatively {\it
consistent} with the expectations for radial flow of an in-medium {\it
hadronic} source (here $\pi^{+}\pi^{-} \rightarrow \rho$) decaying
into lepton pairs. The absolute values of $T_{eff}$ are well below the
hadron line defined by the vacuum $\rho$, as expected, and the other
hadrons freeze-out earlier, due to their smaller coupling to the
pions. It is interesting to note that the large difference of $>$50
MeV in $T_{eff}$ between the vacuum $\rho$ and the $\omega$ (same
mass) quantitatively disappears for the lowest peripheral
``$pp$-like'' selection $dN_{ch}/d\eta$$<$10, as visible in the right
panel of Fig.~\ref{fig5}.

The sudden decline of $T_{eff}$ at masses $>$1 GeV is {\it the other}
most remarkable feature in Fig.~\ref{fig7}. {\it If} the rise is due
to flow, the sudden transition to a seeming low-flow scenario is hard
to reconcile with emission sources which continue to be of dominantly
hadronic origin in this region. A more natural explanation would then
be a transition to a qualitatively different source, implying
dominantly early, i.e. {\it partonic} processes like $q\bar{q}
\rightarrow \mu^{+}\mu^{-}$ for which flow has not yet built
up~\cite{Ruppert:2006hf}. While still
controversial~\cite{vanHees:2006ng}, this may well be the first direct
hint for thermal radiation of partonic origin, breaking hadron-parton
duality for the {\it yield} description in the mass domain.

\section{Future: the options}
\label{future}

There are altogether four options in sight which will have the chance
to advance the field further: at RHIC, LHC, SPS and FAIR. At RHIC,
PHENIX is the only experiment capable to measure lepton pairs both in
the LMR and the IMR region. First results on electron pairs, showing
an excess, have already appeared~\cite{Kozlov:2006tx,Afanasiev:2007xw},
but proper tools for the rejection of low-mass Dalitz and conversion
pairs have not yet been available, and the present data quality is
therefore severely limited by S/B ratios of $<$ 1/1000. A
second-generation upgrade at PHENIX will soon become available to cope
with rejection, relying on a ``hadron-blind-detector'' (a RICH) within
a field-free region including the interaction
vertex~\cite{Kozlov:2006tx}; the required magnet geometry, based on a
double coil arrangement, was proposed in the course of the CERES
R\&D~\cite{Glassel:1987kt}. At the LHC, the LMR region is probably
unaccessible, due to prohibitive charged particle multiplicity
densities, but the (even extended) IMR region is covered both by ALICE
and CMS and may lead to exciting new results. On top, both colliders
offer an excellent chance for a revisit of $pp$ collisions also in the
LMR region, made attractive through record-like Bjorken energy
densities. At the SPS, a continuation of NA60 is conceivable, both for
Pb-beams and for lower beam energies. The existence proof for success
exists; the rest is a (possibly unsurmountable) mixture of elements
which are beyond pure science. At FAIR, finally, a multi-purpose
detector (CBS) is in an early stage of development, meant to also
cover lepton pairs. In all probability, the time horizon is here
beyond a decade.

\section{Conclusions}
\label{conclusions}

Since QM 1982 in Bielefeld, 25 years have passed (a whole generation),
and only now initial central issues seem to be clearing up. The first
accurate measurement of the $\rho$ spectral function in the hot and
dense environment of the QCD phase transition exists, but a solid
theoretical link to chiral restoration is still missing. The first
hint for thermal radiation of mostly partonic origin exists, but it
surely requires theoretical consolidation. The field takes a long
breath indeed...

% \appendix

% \section{}
% \label{}

\end{document}